\documentclass[twocolumn,showpacs,preprintnumbers,amsmath,amssymb]{revtex4}
\usepackage{tabularx,graphicx}

\usepackage{color}
\usepackage{hyperref}
\hypersetup{
    colorlinks=true,
    linkcolor=blue,
    filecolor=blue,      
    urlcolor=blue,
}

\begin{document}

\newcommand{\beq}{\begin{equation}}
\newcommand{\eeq}{\end{equation}}
\newcommand{\beqn}{\begin{eqnarray}}
\newcommand{\eeqn}{\end{eqnarray}}
\newcommand{\bmath}{\begin{subequations}}
\newcommand{\emath}{\end{subequations}}
\newcommand{\bra}[1]{\langle #1|}
\newcommand{\ket}[1]{|#1\rangle}

\title{Inconsistency of the conventional theory of superconductivity}
\author{J. E. Hirsch }
\address{Department of Physics, University of California, San Diego,
La Jolla, CA 92093-0319}

\begin{abstract} 
 In a process where  the temperature of a type I superconductor in a magnetic field changes, the conventional theory of superconductivity predicts that Joule heat is generated and that the final state is
independent of the speed of the process. I show that these two predictions cannot be
simultaneously  reconciled
with the laws of thermodynamics. I propose a resolution of this paradox.
 \end{abstract}
\maketitle

           \begin{figure} [t]
 \resizebox{7.5cm}{!}{\includegraphics[width=6cm]{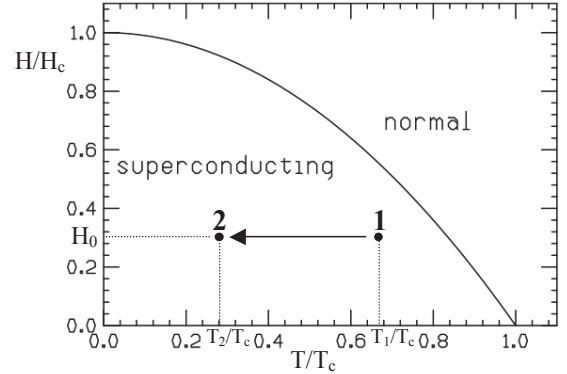}} 
 \caption { Critical magnetic field versus temperature for a type I superconductor. We will consider the process where a system evolves from point 1 to point 2 along
 the direction of the arrow.  }
 \label{figure1}
 \end{figure} 
 
           \begin{figure} 
 \resizebox{8.4cm}{!}{\includegraphics[width=6cm]{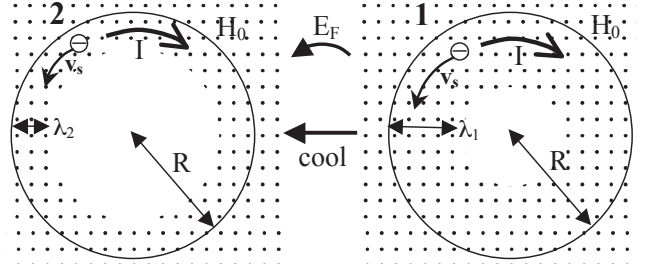}} 
 \caption { Cylindrical superconductor seen from the top. The right (left) panel indicates the system in the state 1 (2) of Fig. 1. 
 The dots indicate magnetic field $H_0$ coming out of the paper. The same current $I$ flows
 in both states.  The Faraday electric field $E_F$ generated during the process points counterclockwise.}
 \label{figure1}
 \end{figure} 

\section{introduction}

Within the conventional London-BCS theory of superconductivity \cite{tinkham}, the state of a simply connected superconductor in an external magnetic field is independent
of how the system reached that state. The theory also predicts that when an electric field
exists in a superconductor at finite temperature, Joule heat is always generated. 
Here I point out that for a type I superconductor in the presence of a magnetic field,
those two assumptions lead to a contradiction with the laws of thermodynamics in a
process where the temperature is changed  below $T_c$.
Consequently, one of the two assumptions must be incorrect in that situation. I propose it is the
second one, and that the alternative theory of hole superconductivity offers a
possible resolution of this paradox.

 \section{the process}

Figure 1 shows the phase diagram of a type I superconductor in a magnetic field $H$ \cite{tinkham}. We consider the process where a cylindrical superconductor 
is cooled from state 1 to state 2 shown in Fig. 1, in the presence of an applied field $H_0$. The inconsistency
also arises if we consider heating instead.
The magnetic field of a long cylinder of radius $R$ and London penetration depth $\lambda_L$ in a magnetic field $H_0$ parallel to its axis is \cite{laue}
\beq
\vec{B}(r)=H_0\frac{J_0(ir/\lambda_L)}{J_0(iR/\lambda_L)} \hat{z}
\eeq
where $J_0$ is the Bessel function of order $0$  and   $\hat{z}$ is along the cylinder axis. To lowest order in $\lambda_L/R$,  
\beq
\vec{B}(r)=H_0 e^{(r-R)/\lambda_L} \hat{z} .
\eeq
The London penetration depth is a decreasing function of temperature, hence a decreasing function of time  in the process of cooling. In the process shown in Fig. 1, the London
penetration depth changes from $\lambda_1$ to $\lambda_2<\lambda_1$ when the temperature is lowered from $T_1$ to $T_2$.
Figure 2 shows the superconductor as seen from the top, with the dots
indicating magnetic field pointing out of the paper.  

\section{Faraday electric field} 

The magnetic field near the surface  is changing in this process, therefore a Faraday electric field is generated. We assume cylindrical symmetry throughout the process.
The electric field at radius $r$ at time $t$  is determined by the equation
\beq
\oint \vec{E}(r,t)\cdot \vec{d\ell}=-\frac{1}{c}\frac{\partial}{\partial t}\int_{r'<r} \vec{B}(r',t)\cdot \vec{dS}
\eeq
which yields
\beq
\vec{E}(r,t)=
-\frac{H_0}{c}(1+\frac{R-r}{\lambda_L})e^{(r-R)/\lambda_L(t)}\frac{\partial \lambda_L}{\partial t}\hat{\theta} .
\eeq
The electric field points counterclockwise.

At any given temperature there are both superfluid and normal electrons, of density $n_s$ and $n_n$, with $n_s+n_n=n$ constant in time, in a two-fluid description \cite{tinkham}. Similarly within BCS theory there is the superfluid and Bogoliubov quasiparticles at finite
temperature, we will call the latter `normal electrons' \cite{tinkham}.
The Faraday electric field will impart  momentum to these normal electrons during the process, and this momentum will decay to zero through scattering with impurities or phonons \cite{tinkham}.
These are irreversible processes, that generate Joule heat and entropy \cite{reif}.  The normal current induced by the Faraday electric field is
\beq
j_n(r,t)=\sigma_n(t)E(r,t)
\eeq
with \cite{tinkham}
\beq
\sigma_n(t)=\frac{n_n(t)e^2\tau}{m^*}
\eeq
within a Drude description with relaxation time $\tau$, with $m^*$ the transport effective mass. The energy dissipated per unit time per unit volume is
\beq
\frac{\partial w(r,t)}{\partial t}=\sigma_n(t) E(r,t)^2,
\eeq
and the energy per unit time dissipated over the entire volume is
\beq
 \frac{\partial W(t)}{\partial t} = \int d^3r \frac{\partial w(r,t)}{\partial t} .
 \eeq
If the process extends from time $t=0$ to $t=t_0$ the total Joule heat dissipated is
\beq
Q_J=\int_0^{t_0}  \frac{\partial W(t)}{\partial t} dt
\eeq
and the Joule entropy generated during this process is
\beq
S_J=\int_0^{t_0}  \frac{\partial W}{\partial t} \frac{1}{T(t)}dt
\eeq
where $T(t)$ is the temperature at time $t$. We assume the process is sufficiently slow that $T(t)$ is well defined at all times.

Note that $Q_J$ and $S_J$ depend on the speed of the process. If we assume for simplicity that $\partial \lambda_L/\partial t$ is constant, we have
\beq
\int_0^{t_0} (\frac{\partial \lambda_L}{\partial t})^2 F(\lambda_L(t))dt=\frac{\partial \lambda_L}{\partial t}\int_{\lambda_1}^{\lambda_2} F(\lambda_L)d\lambda_L
\eeq
for any $F$, so $Q_J$ and $S_J$ are directly proportional to $\partial \lambda_L/\partial t$ . 
In addition, $Q_J$ and $S_J$ are proportional to the Drude relaxation time $\tau$, or equivalently to
the normal state conductivity.

\section{thermodynamics}
 
 We consider the situation shown in Fig. 3. The system is our superconductor with phase diagram given in figure 1, with applied magnetic field
 $H_0$. 
 The system is initially in thermal equilibrium at temperature $T_1$, with London penetration depth $\lambda_1=\lambda_L(T_1)$.
 
We put it in thermal contact with a heat reservoir at temperature $T_2<T_1$ through a  wall  with thermal conductivity
 $\kappa$. Heat will flow and eventually the system will reach temperature $T_2$ and be in thermal equilibrium with the heat
 reservoir. We assume the entire assembly  is thermally and mechanically insulated from its environment.
 The magnetic field originates in external permanent magnets, no work is performed on those magnets during the process. 
 We also assume the process is sufficiently slow that no electromagnetic radiation is generated.  
 Under these conditions, 
 {\it the initial and final states of BOTH  the system AND the reservoir are uniquely determined}.

          \begin{figure} [t]
 \resizebox{7.5cm}{!}{\includegraphics[width=6cm]{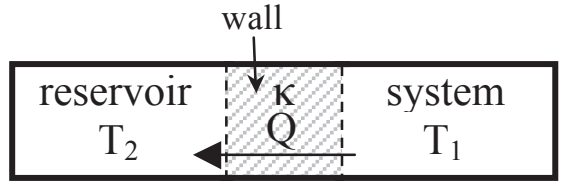}} 
 \caption { The system (superconductor in a magnetic field) at initial  temperature $T_1$ is connected to a heat reservoir at
 temperature $T_2<T_1$ through a wall  of thermal conductivity $\kappa$. The entire assembly is thermally and 
 mechanically insulated from its environment.    }
 \label{figure1}
 \end{figure}

 Given the initial and final states, we can compute various thermodynamic quantities. The total heat $Q$ transferred from the system to
 the reservoir during the process is
 \beq
 Q=\int_{T_2}^{T_1} dT C(T)
 \eeq
 where $C(T)$ is the equilibrium  heat capacity of our system. The change in entropy of the system in this process is
 \beq
 \Delta S=S(T_2)-S(T_1)=\ \int_{T_1}^{T_2} dT  \frac{C(T)}{T}
 \eeq 
 and is of course negative since $T_2<T_1$. The   change in entropy of the universe in this process is
 \beq
 \Delta S_{univ}=\frac{Q}{T_2}+\Delta S
 \eeq
and  is of course positive since we are dealing with an irreversible process, heat conduction between systems at different temperatures.
The quantities $Q$ and $\Delta S_{univ}$ depend $only$ on the initial and final states of the process, $not$ on the
speed at which the process happens.

The Joule heat $Q_J$ and associated entropy $S_J$ discussed in the previous section depend on the speed of the process,
which   will depend principally on the thermal conductivity of the heat conductor,
$\kappa$, connecting the system and the heat reservoir. It would appear that the existence of Joule heat violates both the first and second law of thermodynamics. 

In the next section we will show that even if it may be possible to `save' the first law by some contrived assumption,
the second law is necessarily violated.

\section{the inconsistency}
Let us consider a small  step in the process, starting with the system at temperature $T$, where the system supplies heat 
$\Delta Q$ to the reservoir which is at temperature $T_2<T-\Delta T$. The system will change its temperature from $T$ to $T-\Delta T$. Assume we connect and disconnect the thermal connection between system and reservoir at the
beginning and the end of this step, and wait at the end until equilibrium has been attained. Consider two different ways to do this step:

(a) Infinitely slowly

(b) In a finite amount of time, $\Delta t$.

\noindent According to the previous discussion, for (b) finite  Joule heat will be generated in the
system during this process.

First, let's realize that   the change in temperature of the system, $\Delta T$,    has to be the same for (a) and (b). The reason is, the 
energy transferred to the reservoir was $\Delta Q$ for both (a) and (b), and the final state and hence final temperature of the system is uniquely
determined by its energy, which is the same for (a) and (b), namely (initial energy -$\Delta Q$). Note also that
the energy in the electromagnetic field is also the same in the final state of processes (a) and (b).
We discuss the electromagnetic field in  detail elsewhere \cite{inconsistencylong}, it is not necessary to include it for this argument.
The final state of the reservoir is also unique, depending only on the amount of heat $\Delta Q$ supplied to it, and independent
of the speed at which that heat was supplied to it.

For process (a), we have 
\beq
\Delta Q=C(T) \Delta T
\eeq
where $C(T)$ is the equilibrium heat capacity of the system. 
For process (b), assume Joule heat $\Delta Q_J$ is generated. If we were to assume that the heat capacity of the system is still given by $C(T)$, it would be impossible for the system to reach the same final temperature $T-\Delta T$ upon transferring
energy $\Delta Q$ to the reservoir, instead it
would reach final temperature $T-\Delta T'$, with $\Delta T'$ satisfying
\bmath
\beq
\Delta Q=C(T) \Delta T'+\Delta Q_J  
\eeq
But, as argued in the above paragraph, the system reaches the same temperature $T-\Delta T$ in processes (a) and (b).
Or alternatively, if the heat capacity doesn't change and the system in process (b) changes its temperature by the same amount
$\Delta T$ as in process (a), it would have to 
transfer a different amount of heat to the reservoir
\beq
\Delta Q'=C(T) \Delta T+\Delta Q_J  
\eeq
\emath
instead of $\Delta Q$, again in contradiction with the assumptions.

So let us instead assume  that the heat capacity of the system is
different than the equilibrium one when the process occurs at a finite rate and  involves Joule heat, let's call it $C_r(T)<C(T)$. We will  then have for process (b)
\beq
\Delta Q=C_r(T) \Delta T+\Delta Q_J  
\eeq
transferred from the system to the reservoir, the same as in process (a), with the same change in temperature of the system.
So under this assumption the first law of thermodynamics is not violated, energy is conserved. However, let's consider the change in entropy of the universe. In process (a) it is
\beq
\Delta S_{univ}^{(a)}=\frac{\Delta Q}{T_2}-\frac{\Delta Q}{T}  +O((\Delta T)^2).
\eeq
In process (b), Joule heat $\Delta Q_J$ is generated. 
Quantitatively, we obtain from Eqs. (4), (7) and (8)
\beq
\frac{\partial W}{\partial t}=\sigma_n(\frac{\partial \lambda_L}{\partial t})^2\frac{H_0^2}{c^2}\frac{\pi h R\lambda_L(t)}
{2}
\eeq
and $\Delta Q_J=\int(\partial W/\partial t)dt$ is given, using that
\beq 
  \frac{\partial \lambda_L}{\partial t}=\  \frac{\partial \lambda_L}{\partial T}  \frac{\partial T}{\partial t}
\eeq
by
 \beq
\Delta Q_J= \Delta T \frac{\partial T}{\partial t} \frac{\lambda_L(T)}{2R}\sigma_n(\frac{\partial \lambda_L}{\partial T})^2 \frac{H_0^2}{c^2} V 
\eeq
with
\beq
\frac{\partial T}{\partial t}=\frac{\kappa A}{C(T)}\frac{T-T_2}{d} 
\eeq
and $V=\pi R^2 h$ is  the volume of the cylinder, $d$ the thickness of the wall connecting the reservoir and system and
$A$ its area, and $\sigma_n$ as given by Eq. (6) with $n_n(T)$.

Therefore, in process (b) entropy increases for two reasons. First, generation of Joule heat generates entropy:
\beq
\Delta S_J =  \frac{\Delta Q_J}{T} + O((\Delta T)^2)  .
\eeq
Note that $\Delta Q_J$ and hence $\Delta S_J$  is
$O(\Delta T)$ and not $O((\Delta T)^2)$.
Second, the transfer of the heat $\Delta Q$ from the system to the reservoir generates at least as much entropy as given 
by eq. (18), which is also $O(\Delta T)$. The reason we say `at least' is because the transfer of heat $\Delta Q$ out of
the system lowers its entropy by $-\Delta Q/T$ {\it or less}, by Clausius inequality, if the process takes a finite time. Therefore, the change in entropy of the universe in process (b) is
\beq
\Delta S_{univ}^{(b)} \geq \Delta S_{univ}^{(a)} + \Delta S_J >  \Delta S_{univ}^{(a)}.
\eeq
However, entropy is a function of state. Therefore, the second law of thermodynamics is violated
by Eq. (24).

We can also consider the reverse process, heating the superconductor below $T_c$, as shown in Fig. 4, where the inconsistency
may be even clearer.
          \begin{figure} [t]
 \resizebox{7.5cm}{!}{\includegraphics[width=6cm]{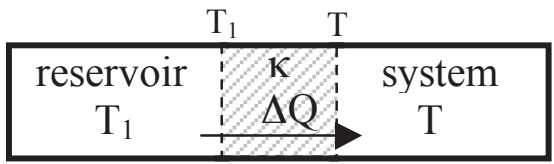}} 
 \caption { Heating: the reservoir is at temperature $T_1$, the system is at temperature $T<T_1$. 
 An amount of heat $\Delta Q$ flows from the reservoir to the system in a time $t_0$ that is
 inversely proportional to $\kappa$ and to $(T_1-T)$.
    }
 \label{figure1}
 \end{figure} 
With the system at temperature $T<T_1$, with $T_1$ the temperature of the reservoir, 
an amount of heat $\Delta Q$ will flow from the reservoir to the system and raise the temperature of the system by
 $\Delta T=\Delta Q/C(T)$. The change in entropy of the universe in this process if it occurs infinitely slowly
 (i.e. if $\kappa\rightarrow 0$)  is
\beq
\Delta S_{univ}=-\frac{\Delta Q}{T_1}+\frac{\Delta Q}{T} + O((\Delta T)^2)  .
\eeq
The time the process actually  takes for finite $\kappa$  is finite,  inversely proportional to $\kappa$ and to $T_1-T$.   The Joule heat generated $\Delta Q_J$
is given by Eq. (21) with $(T_1-T)$ replacing $(T-T_2)$ in Eq. (22). 
We  have to assume that through an unspecified process, which itself may violate the second law,  a  part of the absorbed
$\Delta Q$, in an amount that depends on $\kappa$,  is used up in providing the work  that
propels the normal current, that decays by generation of  $\Delta Q_J$. 
In any event, Joule entropy  
\beq
\Delta S_{extra}=\frac{\Delta Q_J}{T}+ O((\Delta T)^2) 
\eeq
will be generated by the decay of the normal current
that is added to the entropy Eq. (25), violating the second law.

\section{is there a possible resolution within the conventional theory?}
It has been suggested that  one crucial flaw (1) in  my  argument may be 
 the implicit assumption that the sample at intermediate times can be characterized by a uniform temperature T \cite{halperin2}. Another  crucial flaw (2) may be that  even assigning a well-defined temperature to the region where normal current is flowing
may not be possible \cite{halperin2}.  Another   crucial  flaw (3)  may be  that the relaxation time may be a function of momentum
in the superconducting state rendering Eq. (5) invalid \cite{leggett}. Another  crucial flaw (4) may be that assuming that  the only dissipative mechanism in the problem is the Joule heating may render my conclusion invalid \cite{leggett}. 
Another crucial flaw (5) may be that within the conventional theory of superconductivity no electric field can exist in the superconductor unless the current exceeds
a critical value \cite{refereeb}. Another crucial flaw (6) may be that the final state of the reservoir is not unique because
the reservoir is an infinite system \cite{refereea}.
In the following I address those suggestions.

Regarding (3) and (4), I argue that even if those suggestions are valid they would not invalidate my argument. 
If there is another dissipative mechanism in the problem besides Joule heating (and in fact I believe there is within the conventional
theory \cite{entropy}) it would only make the inconsistency worse, since all I need is that there is some dissipation for the inconsistency to arise.
Regarding Eq. (5), even if it needs to be replaced by a more complicated expression that would take into account a relaxation time
that is a function of momentum and/or a non-local generalization of Ohm's law, it would not change the fact that it gives rise to
dissipation. Furthermore, as discussed by Tinkham \cite{tinkham}, Sect. 2.5, a two-fluid approximation with a normal conductivity
given by Eq. (5) ``is the standard working approximation for understanding electrical losses in superconductors'' in situations with 
ac currents or applied electromagnetic fields, and there is no reason to expect within the conventional theory that the same would 
not apply to the situation considered here.

Let us consider the  objection  (1), that the sample may not be at a uniform temperature. First, it should be realized that 
the speed at which temperature equilibrates depends on another variable not included in the argument, namely the thermal conductivity of the sample,
that is at our disposal. We may simply  assume we have a sample with sufficiently high thermal conductivity that it homogenizes the
temperature on a timescale much shorter than all other timescales in the problem.

Still, let us assume that for some unknown reason this does not happen. Considering the heating process of Fig. 4, let us assume the surface
layer heats up and becomes hotter than the bulk, and becomes a `subsystem' at  temperature $T_h=T+\delta T$. One might argue that part of the incoming heat 
$\Delta Q$ provides energy to drive the Joule current in this subsystem at temperature $T_h$, and the resulting Joule heat is dumped into 
the bulk at the lower temperature $T$ as in a `heat engine', thus not violating the second law. The heat coming into the subsystem at
temperature $T_h$ would raise its entropy less
than if its temperature was $T$, and this difference may account for the extra Joule entropy generated.

To counter this argument, we may simply assume that the sample is not heated from the surface but from the interior. Assume the sample is
a hollow cylinder, with no magnetic field in the interior nor in the hollow cavity, so that supercurrent only flows near the outer  surface
as before. The heat $\Delta Q$ is added through the inner surface, so there is no mechanism for the outer surface layer to heat up beyond the bulk before
generating Joule heat.
As the heat $\Delta Q$ is coming in, the London penetration depth will increase, with a corresponding change in magnetic flux and
associated Faraday field generated, and 
 the generated Joule heat $\Delta Q_J$ will generate Joule entropy $\Delta Q_J/T$ thus violating the second law.

Regarding the suggestion (2)  that assigning a well-defined temperature to the region where normal current is flowing may not be possible,
we argue  that even if so it does not eliminate the inconsistency. Particularly in the scenario described in the preceding
paragraph. Furthermore one has to keep in mind that the thickness of the region where the current flows could even be a significant
fraction of the volume of the system, at temperatures sufficiently close to $T_c$ and with sufficiently small magnetic fields,
and if the thermal conductivity of the sample is large and the process not very fast there is no reason to assume
that a large portion of the system would have an undefined temperature. 

Regarding the suggestion (5) that within the conventional theory of superconductivity no electric field can exist
in a superconductor that is smaller than a critical value because it is shorted by the condensate \cite{refereeb},
the suggestion is simply wrong, and reveals a deep lack of understanding of the conventional theory of
superconductivity. Quoting from  Tinkham chapter 2 \cite{tinkham}, {\it ``electric field also acts on the so-called
``normal'' electrons (really thermal excitations from the superconducting ground state,
as we shall see in Chap. 3), which scatter from impurities, and can be described by
Ohm's law.''}

Regarding the suggestion (6) that the final state of the reservoir is not unique because
the reservoir is an infinite system \cite{refereea}, the suggestion is also  wrong and reveals a deep lack of
understanding of thermodynamics. It is easy to see that in the scenario discuss in this paper the final state of the
reservoir is unique whether the reservoir if infinite or finite.  The system starts at temperature $T_1$, the (finite) ``reservoir" starts at temperature $T_2<T_1$, when they have reached thermal equilibrium they will both attain temperature $T_3$, with $T_2<T_3<T_1$. If the ``reservoirÓ is large, 
$T_3$  will be very close to $T_2$, if not it will not, but it doesn't matter. The key point is that the value of $T_3$  cannot depend on whether Joule heat was generated or not, by conservation of energy.
To prove this we just have to remember that energy is a function of state. The energy of the system at temperatures $T_1$ and $T_3$ are fixed, so are the energies of the ``reservoir" at temperatures $T_2$ and  $T_3$. The ``system plus ''reservoir"" is the universe, there is nothing else. So by conservation of energy 
\beq
E_{sys}(T_1)+E_{res}(T_2)=E_{sys}(T_3)+E_{res}(T_3)
\eeq
If, by having the process go at different speed, with different Joule heat generated, the system plus reservoir would attain an equilibrium temperature $T_4$, we would have by conservation of energy
\beq
E_{sys}(T_1)+E_{res}(T_2)=E_{sys}(T_4)+E_{res}(T_4)
\eeq
Therefore combining (27) and (28),
\beq
E_{sys}(T_3)+E_{res}(T_3)=E_{sys}(T_4)+E_{res}(T_4)
\eeq
hence from (29)
\beq
E_{sys}(T_3) - E_{sys}(T_4) = E_{res}(T_4) - E_{res}(T_3)
\eeq
If $T_3$ is not identical to $T_4$, this equation ((30)) implies that either the system or the ÔreservoirÕ have a negative heat capacity. 
I.e. for example, if  $T_3>T_4$ and the left side of (30) is positive, the right side is positive hence $E_{res}(T_3) - E_{res}(T_4)$  is negative, hence the `reservoir' has negative heat capacity. But thermodynamic systems with negative heat capacity can't exist.
Therefore, $T_3=T_4$. Therefore,  the system plus the (finite) `reservoir' have to reach a unique final equilibrium temperature, independent of how much Joule heat is generated in the process. Therefore, all the arguments in this paper apply to the system plus ÔreservoirÕ reaching a unique final temperature $T_3$ with   $T_2<T_3<T_1$. 

Referees of this paper and ref. \cite{leggett} have also raised the question whether  an inconsistency as pointed out here would also
arise in magnetic systems, in particular in ferromagnets. When the temperature is changed and the magnetization
changes, a normal current would be generated by Faraday's law and dissipation would occur that may also be expected
to depend on the speed of the cooling process. 

I have not studied these issues for the case of magnetic systems, so I don't know for sure but  strongly suspect  that the same problem doesn't arise there. There are important differences between ferromagnetism and superconductivity. (i) There is hysteresis in ferromagnets, there is none in type I superconductors; (ii) The ferromagnetic transition is 2nd order, the superconducting one (type I in a magnetic field) is first order; (iii) The Clausius-Clapeyron equation applies to superconductors, not to ferromagnets. So the superconducting transition is more analogous to the water-ice or water-vapor transition than to the ferromagnetic transition; (iv) in magnetic systems there are necessarily other dissipation mechanisms at play: if a magnetic field is applied to a magnetic moment not parallel to the field, the magnetic moment will precess around the magnetic field direction, and its component along the field will not change, hence the magnetization will not change, unless there is a damping mechanism, which will generate entropy. So it is possible that the Joule heating from the Faraday field and this other dissipative mechanism combine to give a total change in entropy that depends only on initial and final states and not on the speed of the process.

\textbf{\textit{}}

In this connection I would also like to quote from Fritz London's 1950 book on superconductivity \cite{londonbook} (emphasis added):
{\it ``So far all attempts to develop a molecular theory of superconductivity
have taken it more or less for granted that it is necessary to assume a
great number of different equilibrium states corresponding to different
spontaneous currents differing in direction and in strength. Apparently
this idea originated from a  \textbf{\textit{quite unwarranted analogy with ferromagnetism.
In the case of ferromagnetism it is in fact possible to
construct a separate state for every orientation of the spontaneous 
magnetic moment.
However, for superconductivity such a construction is of no avail.}} If the various supercurrents really were to correspond to
a continuum of different quantum states it would seem extremely hard
to understand how a supercurrent could resist so many temptations to
dissipate into the other states. Moreover, in this case,      \textbf{\textit{we should expect
to find hysteresis whenever a change of direction or of strength of a
supercurrent is produced, say by changing the direction or the strength
of an applied magnetic field - something similar to ferromagnetic hysteresis.
Nothing of this kind has ever been observed with superconductors,}}
and indeed any hysteresis would be quite incompatible with
all evidence we have concerning the peculiar mobility by which the
supercurrents must adjust themselves to the slightest changes of an
applied external field in order to maintain B = 0... 
The electrodynamics of the superconductor leads to an entirely
different concept. In an isolated, simply connected superconductor and
for a given applied magnetic field, there is just one stable current distribution,
the direction and strength of the current everywhere being
determined by the direction and strength of the external field ...We also saw that for a given external field, 
\textbf{\textit{in contrast to
ferromagnetism, an isolated, simply or multiply connected superconductor
can be assumed to have only one true equilibrium state''.}}}

Finally, in connection with the many comments I have received from referees pointing to the many successes of 
the conventional theory of superconductivity, I would simply like to
point out that it is sufficient
that there is one situation where the inconsistency pointed out in this paper clearly exists to validate our argument. Achilles' heel doesn't have to be more than
a tiny fraction of the entire body area.
 
  \section{discussion}
  The Faraday electric field is a consequence of Maxwell's equations and is unavoidable. The fact that
  electric fields in superconductors give rise to dissipation is well known from experiments with ac currents or electromagnetic waves incident on superconductors  \cite{tinkhamac}, and is predicted by BCS theory \cite{tinkhamac}. So how can this inconsistency be resolved?
  
If the processes occurs always infinitely slowly, the Joule heat and associated entropy go to zero and the 
  inconsistency is resolved. However, there is no mechanism to make these processes proceed infinitely slowly if
  $\kappa$ is large. Furthermore, we know from experiments that superconductors in magnetic fields can be cooled or heated and
  reach equilibrium states at the new temperatures   in finite time.
  
  Another way to resolve this inconsistency would be to assume that the final state depends on the process. Neither I nor
(I suspect) anybody else is willing to go back to that notion, that was discarded in 1933. The contrary notion  is an integral part of
the conventional theory.
  
I  argue that the only other way to resolve this inconsistency is to assume that in the particular situation considered here, where the electric field arises from a change in temperature, the superconductor behaves differently than in other situations with electric fields, 
namely here no normal current is generated and no dissipation takes place.

  That is not predicted by the conventional theory \cite{tinkham}. In addition, within the conventional theory that is impossible, for the following reason. From Ampere's law,
\beq
\oint \vec{B}\cdot \vec{d\ell}=\frac{4\pi}{c}I
\eeq
where  $I$ is the total current, yielding
\beq
I=\frac{c}{4\pi}hH_0 .
\eeq
where $h$ is the height of the cylinder. Therefore, the total current $I$ is independent of temperature. 
However, the Faraday electric field
  transfers   momentum to the supercurrent, as well as to the body as a whole. In order for the current to stay the same, there has to be
  a mechanism for  momentum transfer between electrons and the body as a whole.

  Within the conventional theory of superconductivity, the only way to transfer momentum between electrons
  and ions is through scattering processes involving normal electrons, the same processes that give rise
  to normal resistivity and Joule heat in the normal state \cite{halperin}. If these processes occur at a finite rate
  as in the situation considered here, finite Joule heat and Joule entropy will necessarily be generated. Therefore, the inconsistency cannot be
  resolved within the conventional theory.
 
The only way to transfer momentum between electrons and ions without dissipation other than
infinitely slowly  is if  
electrons  have negative effective mass. If so, an external force acting on the electron gives rise
to acceleration in opposite direction to the force because the difference in momentum is transferred to the body,
without scattering processes and associated dissipation.

This then  implies that  to resolve the inconsistency pointed out in this paper charge  carriers in superconductors have to be holes 
\cite{am} rather than electrons. This is not
required within the conventional theory but is required within the alternative theory of hole superconductivity \cite{holesc}.
We have shown that within that theory there is momentum transfer between electrons and ions 
without dissipation in the normal-superconductor and superconductor-normal transitions
in the presence of a magnetic field
\cite{momentum,revers,entropy}.

\acknowledgements
 The author is grateful to Tony Leggett for motivating him to study this problem and for   extensive 
 and stimulating discussions on this and related problems, 
to Bert  Halperin for extensive and  stimulating discussions on this and related problems that did not reach a conclusion,
and to multiple reviewers for their time and effort spent in reviewing this paper.
Reviewer's reports
can be found in ref. \cite{reviewers}.

    \end{document}